\documentclass[a4paper,10pt]{article}

\usepackage[T1]{fontenc}
\usepackage[utf8]{inputenc}
\usepackage[margin=25mm]{geometry}
\usepackage{booktabs}
\usepackage{array}
\usepackage{hyperref}
\usepackage{tikz}
\usetikzlibrary{positioning,arrows.meta,shapes.geometric}

\newcommand{\tap}{\texttt{tap}}

\title{\tap: A File-Based Protocol for Heterogeneous LLM Agent Collaboration}
\author{Minseo Kim$^\circ$\\
HUA Labs, Seoul, Republic of Korea\\
\texttt{devin@hua-labs.com}}
\date{}

\begin{document}

\maketitle

\begin{abstract}
Existing multi-agent software development systems have proposed many forms of
agent collaboration, including role-based collaboration and automated code
review. However, many systems assume a common runtime, a central conversation
server, or the same API family. Under these assumptions, LLM agents from
different vendors cannot easily exchange messages directly from their own
execution environments while dividing development and review work on a shared
codebase. This paper presents \tap, a file-based collaboration protocol that
allows Claude (Anthropic) and Codex (OpenAI) to collaborate on one codebase
without shared memory or an identical runtime. The core of \tap{} is a
file-first design that preserves markdown files with metadata as original
messages, combines a file inspection path (file communication, Tier 1) with
real-time notification paths for Claude and Codex (real-time communication,
Tier 2), and isolates work through separate git worktrees. Even if real-time
notification fails or a receiver restarts, the message file remains available
and the same content can be inspected again. In a 27-day, 37-generation
self-applied operation where \tap{} was used to develop and review itself, we
collected 209 \tap-related pull requests and 717 operational artifacts. An
analysis of 375 review artifacts showed that the share of reviews recording at
least one defect or requested change was 69.8\% for heterogeneous model pairs
and 53.1\% for homogeneous model pairs. These results show that \tap, which
combines file-based message preservation with real-time notification, operates
in a real production repository, and that combining heterogeneous models and
execution environments can broaden review perspectives. \tap{} is distributed
as the open-source npm package \texttt{@hua-labs/tap} (v0.5.2).
\end{abstract}

\noindent\textbf{Keywords:} multi-agent systems, agent collaboration, software
engineering, large language models

\vspace{6pt}
\noindent\textbf{Archival notice:}
This is an author-prepared English archival translation of the Korean
camera-ready paper accepted to the Korea Computer Congress (KCC) 2026. The
original Korean version is associated with the KCC 2026 proceedings. Copyright
and reuse rights for the proceedings version are governed by the KCC/KIISE
publication agreement. This translation is provided for scholarly archival
access and does not supersede the Korean proceedings version.

\section{Introduction}

Large language model (LLM) based coding agents are being introduced into
practical software development. ChatDev~\cite{chatdev},
MetaGPT~\cite{metagpt}, and AutoGen~\cite{autogen} proposed role-based or
conversation-based multi-agent collaboration, while SWE-agent~\cite{sweagent}
and OpenHands~\cite{openhands} demonstrated LLM-based automated bug fixing.
However, many existing frameworks assume that all agents operate within the
same vendor model family, a common runtime, a central conversation server, or
the same API family. In such structures, authors and reviewers can share
similar error tendencies, and consensus-based strategies may fail to filter
shared errors sufficiently~\cite{vallecillos}. Google A2A~\cite{a2a} proposes a
standard for communication between agents, but it assumes HTTP/SSE-based
network connectivity and therefore addresses a different problem from the
local, repository-centered, file-based collaboration considered in this work.

This paper proposes \tap. \tap{} is a lightweight protocol that uses the file
system as the original message store and combines it with real-time
notification paths specific to each execution environment. This enables agents
from different models to exchange messages directly and to divide development
and review work on the same codebase without shared memory or the same
execution environment. A team of agents across 37 generations used \tap{} for
27 days to develop and review \tap{} itself, leaving 209 pull requests and 717
operational artifacts. Here, a generation is an operational unit in which an
agent team is replaced because of context limits or mission completion and the
next team inherits the previous generation's operational records. The
contributions of this paper are: (1) the design of a file-based collaboration
protocol that supports direct communication between heterogeneous LLM agents
without a common runtime or central conversation server; (2) a report of a
27-day, 37-generation self-applied operation and 717 operational artifacts; and
(3) an analysis of the defect-detection tendencies and limitations observed in
375 review artifacts through generation 37.

\section{Related Work}

ChatDev~\cite{chatdev} models software development as role-based agent
conversation; MetaGPT~\cite{metagpt} structures collaboration through standard
operating procedure artifacts; and AutoGen~\cite{autogen} generalizes
multi-agent conversation frameworks. These systems demonstrate the possibility
of collaboration, but they mainly assume configurations that use a single model
or the same API family. SWE-agent~\cite{sweagent} and
OpenHands~\cite{openhands} demonstrate automated bug fixing on SWE-bench, but
are centered on a single agent or a single execution environment.
Vallecillos-Ruiz et al.~\cite{vallecillos} show that consensus-based strategies
in LLM ensembles can fall into a ``popularity trap'' that amplifies shared
errors, and that diversity-based strategies can mitigate this problem. Google
A2A~\cite{a2a} and AgentMaster~\cite{agentmaster} address interoperability
based on A2A/MCP, but focus respectively on inter-organization integration and
multimodal information retrieval and question answering. In contrast, \tap{}
reports a long-running operational case in which LLM agents from different
vendors used the local file system as a common interface and divided
development and review roles on the same codebase.

\section{\tap{} Architecture}

\tap{} consists of three functions: file messages, per-environment
configuration, and workspace isolation.

\textbf{File messages.}
When an agent calls \texttt{tap\_reply}, a markdown message containing YAML
metadata is first written to \texttt{inbox/}. This file is the original
message, and the subsequent delivery method differs by execution environment.
File communication (Tier 1) is a file inspection path in which MCP tools read
message files in \texttt{inbox/} directly. Real-time communication (Tier 2) is
a real-time notification path that immediately announces newly written messages
through Claude MCP channel notifications and the Codex delivery program's
WebSocket. The two paths have different roles. Real-time communication reduces
conversation latency, while file communication verifies the original file that
serves as the message reference. Therefore, even if real-time communication
fails or a receiver restarts, the message file remains and the same content can
be read again through file-based communication.

\textbf{Per-environment configuration.}
The \texttt{tap add} command inspects the current configuration, creates a
change plan, applies it, and verifies the result. By standardizing this
procedure, it modifies Claude's \texttt{.mcp.json}, Codex's
\texttt{config.toml}, and Gemini's JSON configuration respectively. In other
words, \tap{} does not force one API on every agent; instead, it handles each
execution environment's configuration file through the same procedure.

\textbf{Workspace isolation.}
Each agent works in a separate git worktree, preventing file conflicts. In
addition, PID files, logs, and state files are separated by \texttt{instanceId},
allowing multiple agents of the same kind to run in parallel.

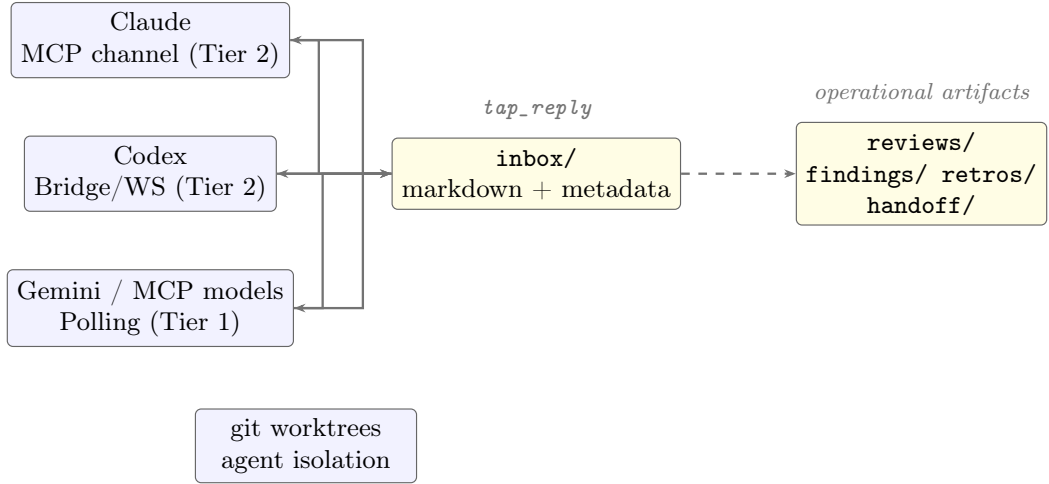
\begin{figure}[t]
\centering
\resizebox{0.86\linewidth}{!}{%
\begin{tikzpicture}[
  node distance=0.45cm and 0.7cm,
  >={Stealth[length=4pt]},
  box/.style={rectangle, rounded corners=2pt, draw=black!60, fill=blue!5,
    minimum width=2.7cm, minimum height=0.65cm, font=\small, align=center},
  file/.style={rectangle, rounded corners=2pt, draw=black!60, fill=yellow!12,
    minimum width=3.0cm, minimum height=0.65cm, font=\small, align=center},
  arrow/.style={->, thick, black!55},
  label/.style={font=\footnotesize\itshape, text=black!55}
]
\node[box] (claude) {Claude\\MCP channel (Tier 2)};
\node[box, below=0.7cm of claude] (codex) {Codex\\Bridge/WS (Tier 2)};
\node[box, below=0.7cm of codex] (polling) {Gemini / MCP models\\Polling (Tier 1)};

\node[file, right=1.4cm of codex] (inbox) {\texttt{inbox/}\\markdown + metadata};
\node[file, right=1.4cm of inbox] (artifacts) {\texttt{reviews/}\\\texttt{findings/ retros/}\\\texttt{handoff/}};
\node[box, below=0.75cm of polling, xshift=1.9cm] (worktrees) {git worktrees\\agent isolation};

\draw[arrow] (claude.east) -- ++(0.35,0) |- (inbox.west);
\draw[arrow] (codex.east) -- (inbox.west);
\draw[arrow] (polling.east) -- ++(0.35,0) |- (inbox.west);
\draw[arrow] (inbox.west) -- ++(-0.35,0) |- (claude.east);
\draw[arrow] (inbox.west) -- (codex.east);
\draw[arrow] (inbox.west) -- ++(-0.35,0) |- (polling.east);
\draw[arrow, dashed] (inbox) -- (artifacts);
\node[label, above=0.1cm of inbox] {\texttt{tap\_reply}};
\node[label, above=0.1cm of artifacts] {operational artifacts};
\end{tikzpicture}%
}
\caption{\tap{} architecture. Messages are first preserved as files, and the
delivery path differs by agent execution environment. File communication (Tier
1) is the file inspection path, while real-time communication (Tier 2) provides
real-time notifications to Claude and Codex.}
\label{fig:arch}
\end{figure}

\textbf{File-first design.}
The file-first structure was chosen because LLM agents from different vendors
differ not only in model family, but also in execution style and communication
interface, making it difficult to assume a common runtime or conversation
channel. In contrast, the local file system is an interface available in most
development environments. \tap{} places message canonicity and auditability in
files, while real-time communication delivers newly written messages
immediately to reduce conversation latency. As a result, even when real-time
communication fails, the message file remains available for reinspection, and
operational records can be used directly for retry and postmortem analysis.

\textbf{Recovery under failure.}
The sender always completes the message-file write before attempting
environment-specific notification. Even when a receiver obtains a real-time
notification, it does not rely on the notification payload itself; instead, it
rereads the original file in \texttt{inbox/} for processing. Therefore, delivery
program restarts, WebSocket disconnections, and lost MCP notifications become
delivery delays rather than message loss. Later, if the file communication path
discovers an unprocessed file, it can receive the same message again. Because
of this structure, the failures in Table~\ref{tab:failures} could be tracked
as protocol-improvement items rather than isolated execution-environment
errors.

\section{Operational Results}

\tap{} was used for 27 days across 37 generations in a real production
repository containing multiple apps and packages, processing 209 \tap-related
pull requests in the process (Table~\ref{tab:metrics}). Real-time communication
was performed with Claude and Codex, while Gemini participated
experimentally through file-based communication.

\begin{table}[t]
\caption{\tap{} operational metrics (commit 563d859c, 2026-04-15 17:00 KST)}
\label{tab:metrics}
\centering
\begin{tabular}{@{}ll@{}}
\toprule
\textbf{Metric} & \textbf{Value} \\
\midrule
Duration & 27 days \\
Agent generations & 37 \\
Merged \tap{} PRs & 209 \\
\tap{} package source & 104 files, 26,404 LOC \\
\tap{} package tests & 76 files, 19,406 LOC \\
Realtime environments & Claude, Codex \\
Generic path & configured MCP polling (exp.) \\
npm & \texttt{@hua-labs/tap} v0.5.2 \\
Artifacts & 717 \\
\bottomrule
\end{tabular}
\end{table}

\textbf{Self-applied operation.}
\tap{} was used to coordinate the development, review, and maintenance of
\tap{} itself: feature additions, defect fixes, code reviews, and operational
record writing were carried out on top of \tap. In this paper, self-hosting is
used in this operational sense and is distinguished from compiler
self-hosting. In early operation, human operators frequently intervened in
message delivery and progress checks. As the real-time notification paths
stabilized, however, the share of direct delivery between agents increased. In
the later phase, a Codex reviewer ran continuously in headless mode on a
separate Linux server. Work instructions and review requests were delivered
through operator-coordinated flows, while major decisions and final approval
remained the operator's responsibility. This configuration provided real
operational conditions in which Claude and Codex collaborated from different
execution environments. Defects discovered during operation did not remain
anecdotal reports; they passed through finding records, mission creation, and
review, leading to protocol changes. Table~\ref{tab:failures} shows examples.

\begin{table}[t]
\caption{Operational failures and protocol fixes}
\label{tab:failures}
\centering
\begin{tabular}{@{}lll@{}}
\toprule
\textbf{Failure} & \textbf{Symptom} & \textbf{Fix} \\
\midrule
Win path prefix & CLI hang & cwd normalize (M233) \\
Bridge log & 0-byte log & per-instance path (M293) \\
Name collision & misdelivery & canonical name (M310) \\
\bottomrule
\end{tabular}
\end{table}

\textbf{Reviews by model pairing.}
From the operational records through generation 37, we cross-checked GitHub PR
lists and inbox review conversations and identified 375 review and re-review
artifacts. The unit of analysis is one review artifact, and the detection rate
is the share of all reviews in which at least one defect or requested change
was recorded. Of the 375 total artifacts, 262 were heterogeneous-model reviews,
and 183 of them (69.8\%) recorded a defect or requested change. There were 113
homogeneous-model reviews, of which 60 (53.1\%) recorded a defect or requested
change. In this operational case, heterogeneous pairings therefore showed a
higher detection rate than homogeneous pairings (Table~\ref{tab:reviews}).

However, 245 samples were Codex reviewers inspecting Claude-authored work, and
later Codex reviewers often ran in a headless environment on a separate Linux
server. Model type, reviewer role, cross-OS effects, and cross-execution
environment effects are therefore entangled. This imbalance is a confounder,
but it can also be interpreted as the result of the operation converging on a
path that enabled rapid feedback and stable unattended review. By manual
reclassification, 12 security or security-adjacent PRs/cases were identified;
6 of these, including an \texttt{execSync} shell injection and a local write
endpoint CSRF, were candidates for runtime security vulnerabilities. These
items are included in Table~\ref{tab:reviews}, but the set also includes
security-feature regressions, quota/rate-limit items, and clean rechecks, so it
is not separated as a security-specific quantitative metric. The results
should therefore be interpreted not as a pure causal effect of model identity,
but as a practical detection tendency observed in an operation that connected
different models and execution environments.

\begin{table}[t]
\caption{Positive review rate for defects or requested changes by
reviewer/author model (Gen 1--37, n=375)}
\label{tab:reviews}
\centering
\begin{tabular}{@{}llll@{}}
\toprule
\textbf{Reviewer} & \textbf{Author} & \textbf{Detected/n} & \textbf{Rate} \\
\midrule
Codex & Claude & 174/245 & 71.0\% \\
Claude & Codex & 9/17 & 52.9\% \\
Claude & Claude & 24/55 & 43.6\% \\
Codex & Codex & 36/58 & 62.1\% \\
Heterogeneous total & & 183/262 & 69.8\% \\
Homogeneous total & & 60/113 & 53.1\% \\
\bottomrule
\end{tabular}
\end{table}

\textbf{Operational records and inter-generational continuity.}
Across 37 generations, 717 operational records accumulated, including
retrospectives, findings, reviews, and handoffs. Individual agents do not
maintain persistent internal state or long-term memory, but the operational
records accumulated as files functioned as external memory through which later
generations could understand work context and decision rationale.

\section{Discussion}

Unlike A2A's~\cite{a2a} HTTP-based inter-organization integration, \tap{} is a
protocol for a small number of agents in the same repository to collaborate
through the local file system and Git workflow. The file system is advantageous
for message preservation and record inspection, while Git is advantageous for
work isolation, change tracking, and forming review units. Remote environments
or inter-organization integration, however, require separate delivery paths.
The difference in detection rates by model pairing suggests that different
models and execution environments can broaden review perspectives, but it
should not be generalized as an inherent superiority of any particular model
pairing.

\textbf{Protocol improvement through self-application.}
Failures and improvement needs encountered during operation were reflected
back into protocol changes or AGENTS instructions through finding records,
retrospectives, missions, and reviews. As this loop repeated across 37
generations, \tap{} operation itself functioned as an evaluation process for
the protocol, and timing issues under concurrent execution as well as platform
defects were naturally exposed (Table~\ref{tab:failures}).

\textbf{Limitations.}
This study has the following limitations. (1) The review data are
observational rather than controlled experimental data, and reviewer assignment
was not random. In particular, later generations include many cases where a
Codex reviewer inspected Claude-authored work, so time effects and reviewer
pairing effects may both be present. (2) Because the unit of analysis is a
review artifact, repeated re-reviews of the same PR may be counted multiple
times; the detection rate should be interpreted as the share of review
artifacts recording a defect or requested change, not as a count of independent
defects. (3) Although \tap{} is a protocol independent of the target code
language, this empirical case was conducted in one operational repository
containing multiple apps and packages and within a single organizational
context; generalization to other organizations, repository structures, and
operational practices requires follow-up validation. (4) Real-time
collaboration was limited to two models, Claude and Codex, while Gemini
participated only experimentally through file inspection. (5) Because \tap{}
itself evolved across 37 generations, early and late operating conditions were
not identical. (6) LLM agent performance can vary with changes in base models,
tool-use ability, and execution environment. (7) The observed difference in
defect detection may reflect model-specific differences, reviewer role
assignment, mission characteristics, and cross-OS or cross-execution
environment effects, which cannot be separated with this dataset alone.
Security-related cases are treated only as qualitative evidence for the
practical value of heterogeneous review, and because the sample is small they
are not included as a separate quantitative comparison by model pairing.

\section{Conclusion}

This paper presented \tap, a file-based protocol for heterogeneous model agent
collaboration, and reported the results of a 27-day, 37-generation
self-applied operation. By combining file-based message preservation with
real-time notification paths, \tap{} showed that Claude and Codex can continue
development and review in one operational repository without a central
coordination server. In the review artifacts through generation 37,
heterogeneous model pairings showed a higher defect detection rate, suggesting
that combining different models and execution environments can broaden review
perspectives. In addition, 717 operational records showed that file-based
records can function as external memory that maintains operational context
across generations even in an agent environment without persistent internal
state. Because these results come from an observational study, future work will
consider balanced comparisons of model pairings, cross-review experiments on
the same PRs, and changes in collaboration quality as models and tools evolve.
\tap{} is distributed as \texttt{@hua-labs/tap} (npm v0.5.2).

\end{document}